\documentclass[letter,twocolumn,amsmath,amssymb,superscriptaddress]{jpsj3}
\usepackage{txfonts}
\usepackage{bm}
\usepackage{color}
\newcommand{\nc}{\newcommand}
\nc{\mb}{\mbox}
\nc{\vs}{\mb{\boldmath$\sigma$}}

\title{Disorder effect on magneto-transport on the surface of a topological insulator}

\author{Idea Matsuzaki\thanks{idea@imr.tohoku.ac.jp} and Kentaro Nomura}
\inst{Institute for Materials Research, Tohoku University, Sendai 980-8577, Japan}

\abst{We study the magneto-transport properties on the disordered surface of a topological insulator attached to a ferromagnet/ferromagnet junction. Since, in the surface Dirac Hamiltonian, an out-of-plane magnetization induces a mass gap, while an in-plane magnetization has a role of the effective vector potential, the mechanism of magneto-transport is different between these two cases. The former is similar to the conventional one in ferromagnetic metals, while the latter is due to the shift of Fermi circles in momentum space. Our numerical calculations show that the magnetoconductance in the in-plane configuration is robust against disorder compared to that in the out-of-plane configuration.}

\begin{document}
\maketitle

Because of many possibilities of application to devices, a magnetoconductance effect has been searched for during the past several decades.
In particular two-dimensional magnetoconductance effect is important to minimize devices.
The localization effect, however, is stronger in two dimension than that of in three dimension and thus the magnetoconductance is fragile against disorder\cite{review_localization}.
Indeed when time-reversal symmetry is broken, all wave functions are localize in two dimension but one exception at the critical point of the quantum Hall (QH) transition\cite{review_QH1,review_QH2,Nomura2008,Nomura2011}.

Although extended wave functions are necessary for the magnetoconductance effect, the QH effect occurs in a strong magnetic field, and thus it is not practical to consider the magnetoconductance in the QH regime. 
Recently, however, the QH effects without an external magnetic field have been realized on the surface of magnetically doped topological insulators (TIs)\cite{review_TI1,review_TI2}.

TIs are new quantum states of matters, which cannot be adiabatically connected to conventional insulators. 
A three-dimensional (3D) TI has a finite gap in the bulk but possesses gapless surface modes described by the two-dimensional (2D) massless Dirac Hamiltonian for simple cases\cite{review_TI1,review_TI2}. 
The surface states are dubbed as the helical surface, in which the spin quantization axis is perpendiculary locked to the momentum by spin-orbit coupling. 

The spin-momentum locking at the surface makes TIs promising for versatile device applications.
At the interface of a ferromagnetic insulator and a topological insulator, a variety of unique magneto-transport phenomena has been theoretically proposed\cite{Burkov2010,Yokoyama2010,Mondal2010,Salehi2011} and experimentally examined\cite{Ando2014,Tian2014}.

In this work we study the disorder effects on the magnetoconductance of topological surface attached to a ferromagnet/ferromagnet junction.
The ferromagnetism on the surface is induced by the exchange interaction ${\bf m}\cdot\vs$, where $\vs=(\sigma_x,\sigma_y,\sigma_z)$ is the Pauli spin matrix of the surface electrons and ${\bf m}=(m_x,m_y,m_z)$ is the exchange field which has the direction of the magnetization and the magnitude of the exchange splitting energy.
The out-of-plane exchange field generates a mass gap in the surface modes. 
When the Fermi level is located slightly above the bottom of the conduction band, by projecting into the conduction band, the surface states can be regarded as fully spin-polarized 2D electrons with conventional parabolic dispersion. 
As in the conventional ferromagnetic metals, the conductance shows a change depending on whether the magnetizations of adjacent ferromagnets are in a parallel or an antiparallel alignment, namely magnetoconductance.
On the other hand, the in-plane exchange field acts as an effective vector potential which shifts the Fermi circles in momentum space. 
A misalignment of the Fermi circles between two regions also gives rise to a magnetoconductance\cite{Yokoyama2010,Mondal2010,Salehi2011}.
With the use of the transfer matrix method\cite{review_transfer_matrix,Bardarson2007,Takane2014}, we calculate magnetoconductance in the out-of-plane and the in-plane magnetization configurations, and compare the disorder dependence of them.
Our result shows a difference between the two cases, that the in-plane magnetoconductance is relatively robust against disorder, compared with the out-of-plane magnetoconductance.

\begin{figure}[b]
\begin{center}
\includegraphics[width=0.4\textwidth]{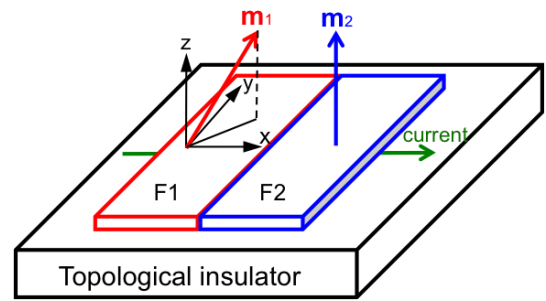}

\caption{Schematic picture of F1/F2 junction on the TI surface\cite{Yokoyama2010}. The ferromagnetism on the surface of a TI is induced due to proximity effect by the ferromagnetic insulators deposited on the surface. The current flows on the surface of a TI.}
\label{setting}

\end{center}
\end{figure}

We consider the surface of a TI which is attached to two ferromagnetic insulators, F1 and F2 (Fig. \ref{setting})\cite{Yokoyama2010}.
The surface electronic states in this system can be described by the 2D Dirac Hamiltonian
\begin{equation}
H=v\left(p_{y}\sigma_{x}-p_{x}\sigma_{y}\right)+\sum_{i=x,y,z}m_{i}\sigma_{i}+U(x,y),
\end{equation}
where $\mathit{v}$ is the velocity of the Dirac fermion,
$\mathit{m_{x}}$, $\mathit{m_{y}}$ and $\mathit{m_{z}}$ are the exchange
fields and $\mathit{U(x,y)}$ is the disorder potential. 
We note the exchage field in F1 as $\mathbf{m}_{1}=(m_{1x},m_{1y},m_{1z})$ and F2 as $\mathbf{m}_{2}=(m_{2x},m_{2y},m_{2z})$. 
We assume the TI's surface which is attached to two ferromagnetic insulators as 2D sheet of length $2L$, where the length of each ferromagnetic insulator is $L$, along longitudinal $\mathit{x}$ direction and width $\mathit{W}$ along transversal $\mathit{y}$ direction. 
We take the aspect ratio $\mathit{W/L}=4, 8,12$. 

We utilize the transfer matrix method\cite{Bardarson2007,review_transfer_matrix,Takane2014} for the 2D Dirac equation
$H\Psi=E\Psi$, where $\Psi(x,y)$ is the two-component (spinor) wave function.
Multiplication of both sides by $i\sigma_{y}$ gives
\begin{equation}
\hbar v\partial_{x}\Psi=\left(v\sigma_{z}p_{y}+i\left(U(x,y)-E\right)\sigma_{y}+m_{x}\sigma_{z}+im_{y}-m_{z}\sigma_{x}\right)\Psi.
\end{equation}
Here we discretize variable $\mathit{x}$ at the $\mathit{N}$ points $\mathit{x_{\mathrm{1}},}x_{2},.$..,$\mathit{x_{N}}$.
The transfer operator $\mathcal{M}$ , defined by $\Psi_{2L}=\mathcal{M}\Psi_{0}$,
is given by the operator product
\begin{equation}
\mathcal{M}=\mathcal{P}_{2L,x_{N}}\mathcal{K}_{N}\mathcal{P}_{x_{N},x_{N-1}}\mathcal{K}_{N-1}...\mathcal{K}_{2}\mathcal{P}_{x_{2},x_{1}}\mathcal{K}_{1}\mathcal{P}_{x_{1},0},
\end{equation}
\begin{equation}
\mathcal{P}_{x_{i+1},x_{i}}=\exp\left[\mathit{\left(\mathrm{1/\hbar}\right)ap_{y}\sigma_{z}}\right].
\end{equation}
The operator $\mathcal{P}$ gives the decay of evanescent waves between two scattering events, described by operator $\mathcal{K}_{n}$, 
\begin{equation}
\mathcal{K_{\mathit{n}}}=\mathcal{V_{\mathit{n}}B_{\mathit{x,n}}B_{\mathit{y,n}}B_{\mathit{z,n}}},
\end{equation}
where
\begin{equation}
\mathcal{V}_{n}=\exp\left[\left(ia/\hbar v\right)(U_{n}-E)\sigma_{y}\right],
\end{equation}
and
\begin{eqnarray}
\mathcal{B}_{x,n}&=&\exp\left[\left(a/\hbar v\right)m_{x}\sigma_{z}\right],
\\
\mathcal{B}_{y,n}&=&\exp\left[\left(ia/\hbar v\right)m_{y}\right],
\\
\mathcal{B}_{z,n}&=&\exp\left[\left(-a/\hbar v\right)m_{z}\sigma_{x}\right],
\end{eqnarray}
where $a$ is the lattice constant.

To calculate the transfer matrix, we represent the operators in the
basis
\begin{equation}
\psi_{k}^{\pm}=\frac{1}{\sqrt{W}}e^{iq_{k}y}|\pm\rangle,\; q_{k}=\frac{2\pi k}{W},\; k=0,\pm1,\pm2,...
\end{equation}
The spinors $|\pm\rangle=2^{-1/2}\begin{pmatrix}1\\
\pm i
\end{pmatrix}$ are eigenvectors of $-\sigma_{y}$.  
By truncating the transverse momenta $\mathit{q_{k}}$ at $|k|=M$, the dimension of the transfer matrix becomes finite. 
The disorder potential $U(x,y)=\sum_{n,m}\gamma_{nm}\delta(x-x_{n})\delta(y-y_{m})$ is given by a collection of isolated impurities distributed uniformly over the scattering region $0<x<2L, 0<y<W$. 
The strengths $\gamma_{nm}$ of the scatterers are uniform in the
interval $\left[-\gamma_{0},\,\gamma_{0}\right]$. The disorder strength
is quantified by the correlator $K_{0}=\frac{1}{(\hbar v)^{2}}\int d\mathbf{r'}\langle U(\mathbf{r})U(\mathbf{r}')\rangle$
which evaluates to $K_{0}=\frac{1}{3}\gamma_{0}^{2}(1/\hbar va)^{2}$, independent of the correlation lengths. 
The disorder strength can be related to the mean free path $\ell$ in Boltzmann transport limit by $\ell=\hbar v/K_{0}E$.
The average conductance $\langle G\rangle$ is obtaibed by sampling some 200-2000 disorder realizations of the impurity potential.
We take $2L/a$ large enough so that the calculation result does not depend on the orders of operator products ($2L/a>20$).
The momentum cutoff $M$ is also large enough in this calculation ($M>20$).

To formulate the scattering problem\cite{method1,method2,method3}, we consider a scattering state $\Psi_{k}$ that has unit incident current from the left ($x=0$) in mode $k$ and zero incident current from the right ($x=2L$). The quantum number $k$ labels transverse modes.
At $x=0$, the sum of incoming and reflected waves given by
\begin{equation}
\label{S_left}
\Psi_{k}^{\mathrm{left}}=\phi_{k}^{+}+\sum_{k'}r_{k'k}\phi^{-}_{k'},
\end{equation}
while the sum of transmitted waves at $x=2L$ is given by
\begin{equation}
\label{S_right}
\Psi_{k}^{\mathrm{right}}=\sum_{k'}t_{k'k}\phi^{+}_{k'}.
\end{equation}
The right moving component in mode $k$ is $\phi_{k}^{+}$ and left moving component is $\phi_{k}^{-}$.
Starting from a mode incident from right, we can similarly obtain the reflection and transmission matrices $r'$ and $t'$, 
which give, with $r$ and $t$, the unitary scattering matrix,
\begin{equation}
S=
\begin{pmatrix}
r & t'\\
t & r'
\end{pmatrix}.
\end{equation}
As a consequence of unitarity, the matrix product $tt^{\dagger}$ and $t't'^{\dagger}$ have the same eigenvalue called transmission eigenvalues.
The conductance $G$ follows from transmission eigenvalues via the Landauer formula
$
G=\frac{e^{2}}{h}\mathrm{Tr}\left[tt^{\dagger}\right]=\frac{e^{2}}{h}\mathrm{Tr}\left[t't'^{\dagger}\right]
$.
 
The information contained in the scattering matrix $S$ can equivalently be represented by transfer matrix $\mathcal{M}$\cite{method1,method2}.
While the scattering matrix relates outgoing waves to incoming waves, the transfer matrix relates waves at the right to wave at the left,
\begin{equation}
\label{M_eq}
\Psi^{\mathrm{right}}=\mathcal{M}\Psi^{\mathrm{left}}.
\end{equation}
We separate the spinor degree of freedom of $\mathcal{M}$ into four blocks
\begin{equation}
\mathcal{M}=
\begin{pmatrix}
\mathcal{M}^{++} & \mathcal{M}^{+-}\\
\mathcal{M}^{-+} & \mathcal{M}^{--}
\end{pmatrix}.
\end{equation}
As one can verify by substitution into Eq. (\ref{M_eq}), and comparison Eq. (\ref{S_left}) and Eq. (\ref{S_right}),
the submatrices $\mathcal{M}^{ss'}\;\left(s,s'=\pm\right)$ are related
to the transmission and reflection matrices by
\begin{eqnarray}
&&r=-\left(\mathcal{M}^{--}\right)^{-1}\mathcal{M}^{-+},\quad
r'=\mathcal{M}^{+-}\left(\mathcal{M}^{--}\right)^{-1}
\\
&& t=\mathcal{M}^{++}-\mathcal{M}^{+-}\left(\mathcal{M}^{--}\right)^{-1}\mathcal{M}^{-+}, \ \ 
t'=\left(\mathcal{M}^{--}\right)^{-1}.
\end{eqnarray}

The repeated multiplication of transfer matrices is unstable because it produces both exponentially growing and exponentially decaying eigenvalues, and the limited numerical accuracy prevents one from retaining both sets of eigenvalues. We resolve this obstacle by converting the transfer matrix into a unitary matrix, which has only eigenvalues of unit absolute value\cite{method1,method2,method3}.

\begin{figure}[t]
\begin{center}
\includegraphics[width=0.5\textwidth]{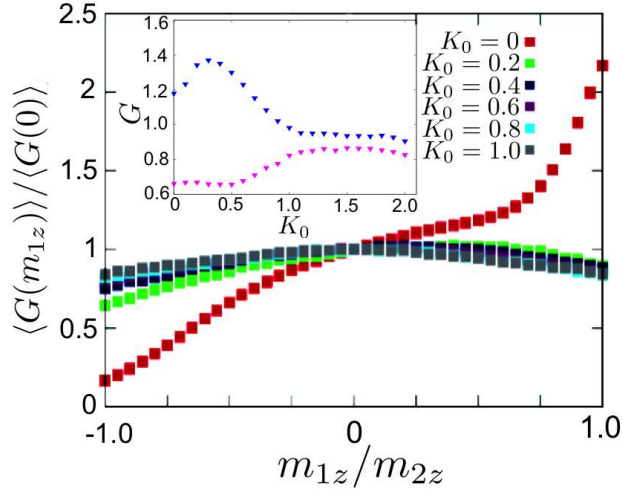}

\caption{The out-of-plane exchange field dependence of normalized conductance for various disorder strength. Here ${\bf m}_1=(0,0,m_{1z})$, ${\bf m}_2=(0,0,m_{2z})$ and $m_{1z}$ is varied from $-m_{2z}$ to $+m_{2z}$ with fixing $m_{2z}=0.2\hbar v/a$. The aspect ratio and the energy are fixed at $W/L=4$ and $E=1.35 m_{2z}$. The inset shows the disorder strength dependence of the conductance for $E=0$ (pink) and $E=0.27\hbar v/a $ (blue).}
\label{out-of-plane}

\end{center}
\end{figure}

This completes the description of our numerical method.
We now turn to the result.
The inset in Fig. \ref{out-of-plane} shows the disorder strength dependence of conductance for $E=0$ and $E=0.27\hbar v/a$ in the absence of exchange field.
The disorder dependence qualitatively changes around $K_{0}\sim 0.3$, indicating the crossover between the ballistic and the diffusive regimes.
Indeed the mean free path is comparable to the system size around this point.
The main panel shows the magnetoconductance in the out-of-plane exchange field configuration.
Here we fix $m_{2z}=0.2\hbar v/a$ and vary $m_{1z}$ from $-m_{2z}$ to $+m_{2z}$.
The Fermi energy $E$ is also fixed at $E=0.27\hbar v/a=1.35 m_{2z}$.
The normalized magnetoconductance is plotted as function of $m_{1z}$ for various disorder strength.

In the clean limit, $K_0=0$, the magnetoconductance is positive for the parallel configuration ($m_{1z}/m_{2z}>0$), while it is negative for the antiparallel configuration ($m_{1z}/m_{2z}<0$).
We see good agreement with analytical results of the out-of-plane field dependence in the continuum model\cite{Yokoyama2010}.

To understand this magnetoconductance behavior, we consider the case where $|m_z|\gg v\hbar k_{F}$, $k_{F}$ being Fermi wave number, and focus on the positive energy band on the surface of a TI.
In this case, the spin degeneracy is lifted and the spin direction is $(-\hbar vk_{y}, \hbar vk_{x},m_{z})$. The energy dispersion approximated as $E=\sqrt{(\hbar vk)^{2}+m_{z}^{2}}\sim \frac{(\hbar vk)^{2}}{2|m_{z}|}+|m_{z}|$, regarded as fully spin polarized Schr\"odinger electrons.
For this reason, the mechanism of this magnetoconductance induced by out-of-plane exchange field corresponds to that in a conventional ferromagnetic metal.
We note the Landau level gaps are negligibly smaller than the energy gap induced by the out-of-plane exchange field\cite{Chen2010}.

Next, we consider the influence of disorder on the surface of a TI. 
Figure \ref{out-of-plane} shows that the $m_{1z}$ dependence of the conductance changes abruptly when weak disorder is introduced. 
The conductance $\langle G(m_{1z})\rangle$ does not become maximum but minimum at $m_{1z}=+m_{2z}$ (parallel configuration), in contrast to the clean limit case. In the disordered case the conductance takes it maximum at $m_{1z}\sim 0$.
At strong disorder, $K_0=1.0$, $m_{1z}$ dependence of the conductance is nearly symmetric around $m_{1z}=0$, indicating that the parallel/antiparallel configuration does not matter.\par

\begin{figure}[t]
\begin{center}
\includegraphics[width=0.5\textwidth]{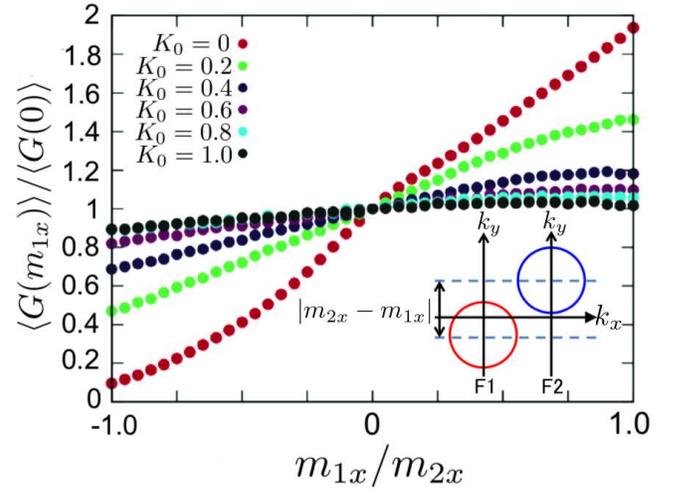}

\caption{The in-plane exchange field dependence of normalized conductance for various disorder strength, ${\bf m}_1=(m_{1x},0,0)$, ${\bf m}_2=(m_{2x},0,0)$ and $m_{1x}$ is varied from $-m_{2x}$ to $+m_{2x}$ with fixing $m_{2x}=0.2\hbar v/a$. The aspect ratio and the energy are fixed, $W/L=4$ and $E=1.35 m_{2x}$.}
\label{in-plane}

\end{center}
\end{figure}

When the exchange field is applied in the $x$ direction, $k_{y}\sigma_{x}$ term in the original Dirac Hamiltonian is replaced by $(k_{y}+m_{x}/\hbar v)\sigma_{x}$ indicating that the Fermi circle is shifted by $-m_{x}/\hbar v$ in momentum space, while the size of Fermi circle remains unchanged.
When $m_{1x}\neq m_{2x}$, the position of the Fermi circles in F1 and F2 are different.
This misalignment of the Fermi circles causes a change of conductance (magnetoconductance) with qualitatively different mechanism from conventional one.
Namely, as $|m_{2x}- m_{1x}|$ increases, the overlap region of the Fermi circles between F1 and F2 is reduced as depicted in the inset in Fig. \ref{in-plane}, and thus the number of evanescent modes increases, therefore the conductance decreases.
The in-plane exchange field induced magnetoconductane is characteristic to the spin-momentum locking and qualitatively different from the conventional magnetoconductance. 

The normalized magnetoconductance in the in-plane configuration, ${\bf m}_{1}=(m_{1x},0,0)$ and ${\bf m}_{2}=(m_{2x},0,0)$, is shown in Fig. \ref{in-plane}.
We fix the exchange field in F2 at $m_{2x}(a/\hbar v)=0.2$, and vary the exchange field $m_{1x}$ from $-m_{2x}$ to $+m_{2x}$.
In the clean limit ($K_{0}=0$), the conductance increases for the parallel configuration ($m_{1x}/m_{2x} >0$), and decreases for the antiparallel configuration ($m_{1x}/m_{2x} <0$).
As mentioned above, the conductance is influenced by the relative positions of the Fermi circles.
The conductance takes maximum when there is no misalignment of Fermi circles between F1 and F2, and decreases with introducing the misalignment.
Again we see good agreement with analytical result in the continuum model\cite{Yokoyama2010}.\par
As Fig. \ref{in-plane} shows, the dependence of magnetoconductance on the in-plane exchange field becomes gradually weak as the disorder strength increases, in contrast to the case of out-of-plane dependence.
Even in the  presence of disorder, the magnetoconductance remains positive for the parallel configuration ($m_{1x}/m_{2x}>0$), while negative for the antiparallel configuration ($m_{1x}/m_{2x}<0$).

\begin{figure}[t]
\begin{center}
\includegraphics[width=0.5\textwidth]{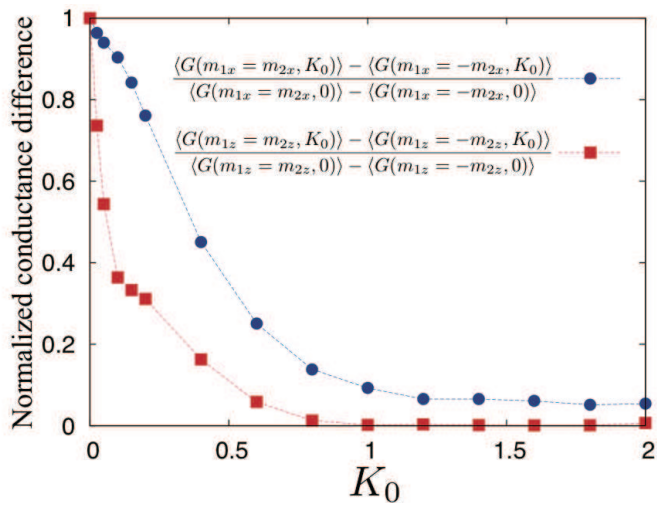}

\caption{The disorder dependence of the normalized conductance difference between parallel and antiparallel exchange field configurations for out-of-plane (red) and in-plane (blue) case.}
\label{disorder}

\end{center}
\end{figure}

To compare disorder dependence of the magenetoconductance with the out-of-plane exchange field and that with the in-plane, we plot the conductance differences $\langle G(m_{1z/x}=m_{2z/x},K_{0})\rangle-\langle G(m_{1z/x}=-m_{2z/x},K_{0})\rangle$ as a function of disorder strength $K_0$ in Fig. \ref{disorder}.
The conductance difference with the out-of-plane field abruptly decreases as the disorder strength, $K_{0}$, increases even in the ballistic regime and vanishes in the diffusive regime.
On the other hand, the conductance difference with the in-plane field decays more slowly and remains finite even in the diffusive regime.
These behaviors are found at all the aspect ratios we examined, $W/L=4, 8,12.$
These results clearly indicate that the magnetoconductance with the in-plane field is robust against disorder while that with the out-of-plane field is fragile.
Since the former is characteristic to the surface states of a topological insulator (strong spin-orbit coupling), it is a great advantage of topological insulator based devices.

In conclusion, we have studied the disorder effect on the magnetoconductance of the ferromagnet/ferromagnet junction on the surface of a TI. 
With the use of the transfer matrix method, we calculated the magnetoconducrance  in both the out-of-plane and the in-plane exchange field configurations.
In the out-of-plane field, the Dirac electrons are regarded as fully spin-polarized Schr\"odinger electrons when the Fermi level is located slightly above the bottom of the conduction band. The mechanism of the magnetoconductance in this regime corresponds to that in a conventional ferromagnetic metal. On the other hand, the in-plane field induced magnetoconducrance is characteristic to the surface of a TI. These two cases show different disorder dependence. 
These results are consistent with the fact that all wave functions are localized in the presence of a mass gap, while in the in-plane fields massless Dirac fermion systems belong to the critical point of the quantum Hall transition, and thus wave functions are extended.\cite{Nomura2008,Nomura2011}
Since the latter is robust against disorder, it is an advantage of TI based devices.

\begin{acknowledgment}
The authors are grateful to K. Kobayashi and J. Barker for helpful arguments.
This work was supported by Grant-in-Aid for Sci- entific Research (No. 15H05854, No. 26107505 and No. 26400308) from the Ministry of Education, Culture, Sports, Science and Technology (MEXT), Japan.
\end{acknowledgment}

\end{document}